\documentclass[useAMS,usenatbib,usegraphicx]{mn2e}
\usepackage{amssymb}

\title[Cosmic-ray pressure in galaxy clusters]{Gamma-ray probe of
cosmic-ray pressure in galaxy clusters and cosmological implications}
\author[S. Ando and D. Nagai]{Shin'ichiro Ando\thanks{E-mail:
ando@caltech.edu} and Daisuke Nagai\thanks{E-mail:
daisuke@caltech.edu}\\%
California Institute of Technology, Mail Code 130-33,
Pasadena, CA 91125, USA}
\begin{document}

\date{Accepted 19 January 2008; submitted 19 December 2007; in original
form 16 May 2007}

\pagerange{\pageref{firstpage}--\pageref{lastpage}} \pubyear{2008}

\maketitle

\label{firstpage}

\begin{abstract}%
Cosmic rays produced in cluster accretion and merger shocks provide
 pressure to the intracluster medium (ICM) and affect the mass
 estimates of galaxy clusters.  Although direct evidence for
 cosmic-ray ions in the ICM is still lacking, they produce
 $\gamma$-ray emission through the decay of neutral pions produced in
 their collisions with ICM nucleons.  We investigate the capability of
 the {\it Gamma-ray Large Area Space Telescope (GLAST)} and imaging
 atmospheric {\v C}erenkov telescopes (IACTs) for constraining the
 cosmic-ray pressure contribution to the ICM.  We show that {\it
 GLAST} can be used to place stringent upper limits, a few per cent
 for individual nearby rich clusters, on the ratio of pressures
 of the cosmic rays and thermal gas.  We further show that it is
 possible to place tight ($\lesssim 10\%$) constraints for distant
 ($z\lesssim 0.25$) clusters in the case of hard spectrum, by
 stacking signals from samples of known clusters. The {\it GLAST} limits
 could be made more precise with the constraint on the cosmic-ray
 spectrum potentially provided by IACTs.  Future $\gamma$-ray
 observations of clusters can constrain the evolution of cosmic-ray
 energy density, which would have important implications for
 cosmological tests with upcoming X-ray and Sunyaev-Zel'dovich effect
 cluster surveys.
\end{abstract}

\section{Introduction}
\label{sec:Introduction}

Clusters of galaxies are potentially powerful observational probes of
dark energy, the largest energy budget in the Universe causing the
cosmic acceleration \cite*[e.g.,][]{Haiman01,Albrecht06}.  Most of the
cosmological applications using clusters rely on the estimates of
their total virial mass---quantity which is difficult to measure
accurately in observations.  Clusters offer a rich variety of
observable properties, such as X-ray luminosity and temperature
\cite*[e.g.,][]{Rosati02}, Sunyaev-Zel'dovich effect
(SZE) flux \citep*[e.g.,][]{Carlstrom02},
gravitational lensing of distant background galaxies
\citep[e.g.,][]{Smith05,Dahle06,Bradac06}, and velocity dispersion
of cluster galaxies \citep[e.g.,][]{Becker07} and proxies for cluster
mass.

One of the most widely used methods for measuring cluster masses
relies on the assumption of hydrostatic equilibrium between
gravitational forces and thermal pressure gradients in the
intracluster medium (ICM) \citep*{Sarazin86,Evrard96}.  Current X-ray
and SZE observations can yield mass of individual clusters very
precisely based on accurate measurements of the density and
temperature profiles
\citep*{Pointecouteau05,Vikhlinin2006,LaRoque06}. However, the accuracy
of the hydrostatic mass estimates is currently limited by nonthermal
pressure provided by cosmic rays, turbulence, and magnetic field in
the ICM \citep*[][and references
therein]{Ensslin1997,Rasia2006,Nagai2007a}.  This nonthermal bias must
be understood and quantified before the requisite mass measurement
accuracy is achieved.  Comparisons with the mass estimates from
gravitational lensing can provide potentially useful limits on this
nonthermal bias \citep[see e.g.,][]{Mahdavi07}.  However, present
observations do not yet constrain the nonthermal pressure in the
regime in which it dramatically affects the calibration of the
hydrostatic mass estimates. If not accounted for, these nonthermal
biases limit the effectiveness of upcoming X-ray and SZE cluster
surveys to accurately measure the expansion history of the Universe.
Detailed investigations of sources of nonthermal pressure in clusters
are thus critical for using clusters of galaxies as precision
cosmological probes.

There is growing observational evidence for the nonthermal activity in
clusters.  For example, radio and hard X-ray observations of clusters
suggest presence of relativistic electrons.  This also implies
presence of relativistic protons produced in the shock that
accelerated these electrons.  However, the signature $\gamma$-ray
emission due to decays of neutral pions produced in the collisions of
cosmic rays with nucleons in the ICM has not been detected.  From
non-detection of $\gamma$-ray emission from clusters with the
Energetic Gamma Ray Experimental Telescope (EGRET) in the GeV band
(\citealt{Reimer2003}; but see also \citealt{Kawasaki2002,
Scharf2002}), constraints have been placed on the fraction of
cosmic-ray pressure in nearby rich clusters at less than $\sim$20\%
\citep[Virgo and Perseus clusters]{Ensslin1997,Miniati2003} and less
than $\sim$30\% \citep[Coma cluster]{Pfrommer2004}.  Similar
constraints are obtained using the Whipple {\v C}erenkov telescope in
the TeV band \citep{Perkins2006}.  These measurements indicate that
the cosmic rays provide relatively minor contribution to the dynamical
support in the ICM \citep[e.g.,][]{Blasi1999}.  However, the current
constraints are too loose for the future cluster-based cosmological
tests.

The next generation of $\gamma$-ray detectors, such as {\it Gamma-ray
Large Area Space Telescope (GLAST)} and imaging atmospheric {\v
C}erenkov telescopes (IACTs), will be able to provide dramatically
improved constraints on the cosmic-ray pressure in clusters, and may
even detect $\gamma$-ray radiation from several rich clusters
\citep[and references therein]{Ando2007}.  The {\it GLAST} satellite,
which is soon to be launched, is equipped with the Large Area
Telescope (LAT) that enables all sky survey with GeV $\gamma$-rays.
Several IACTs are currently working or planned for detecting TeV
$\gamma$-rays, which include HESS, MAGIC, VERITAS, and CANGAROO-III.
Confronting the recent advances in $\gamma$-ray astronomy as well as
growing interests in dark energy studies, in the present paper, we
investigate the sensitivity of these detectors to high-energy
$\gamma$-rays of cosmic-ray origin.

We first show updated sensitivities of {\it GLAST} and IACTs for
nearby rich clusters following \citet{Pfrommer2004}.  In particular,
{\it GLAST} would be able to constrain the cosmic-ray energy density
in such clusters to better than {\it a few per cent} of the thermal
energy density, while IACTs would be useful to constrain the
cosmic-ray spectrum.  We then consider stacking many $\gamma$-ray
images of distant clusters to probe the evolution of cosmic-ray
pressure.  We show that, by stacking many massive clusters, the
upcoming {\it GLAST} measurements will have the statistical power to
constrain the cosmic-ray pressure component to better than $\sim$10\%
of the thermal component for clusters out to $z \lesssim 0.25$.  These
forthcoming measurements will be able to place stringent limits on the
bias in the cluster mass estimates and hence provide important handle
on systematic uncertainties in cosmological constraints from upcoming
X-ray and SZE cluster surveys.

Throughout this paper, we adopt the concordance cosmological model
with cold dark matter and dark energy ($\Lambda$CDM), and use
$\Omega_{\rm m} = 0.3$, $\Omega_{\Lambda} = 0.7$, $H_0 = 100 ~ h$ km
s$^{-1}$ Mpc$^{-1}$ with $h = 0.7$, and $\sigma_8 = 0.8$.

\section{\boldmath{$\gamma$}-ray production due to proton--proton
  collisions}
\label{sec:Gamma-ray production due to proton-proton collisions}

Cosmic-ray protons are injected in the ICM through the shock wave
acceleration, and the momentum distribution follows the power law,
$p_{\rm p}^{-\alpha_{\rm p}}$ with $\alpha_{\rm p} \simeq 2$--3.
These cosmic-ray protons then interact with the surrounding ICM (mostly
nonrelativistic protons), producing neutral and charged pions; the
former decays into two photons ($\pi^0 \to 2\gamma$) while the latter
into electrons, positrons, and neutrinos.  The volume emissivity of the
$\pi^0$-decay $\gamma$-rays (number per volume per unit energy range) at
distance $r$ from the cluster center is given as
\citep[e.g.,][]{Blasi1999b}
\begin{eqnarray}
 q_\gamma (E,r) & = & 2 n_{\rm H}(r) c
  \int_{E_{\pi,{\rm min}}}^{\infty} {\rm d}E_\pi
  \int_{E_{\rm p,th}(E_{\pi})}^{\infty} {\rm d}E_{\rm p}
  \nonumber\\&&{}\times
  \frac{{\rm d}\sigma_{\rm pp}}{{\rm d}E_\pi} (E_\pi,E_{\rm p})
  \frac{n_{\rm p}(E_{\rm p},r)}{\sqrt{E_\pi^2 - m_\pi^2 c^4}},
  \label{eq:emissivity}
\end{eqnarray}
where $m_\pi$ and $E_\pi$ is the mass and energy of the neutral pion,
$E_{\pi,{\rm min}} = E + m_\pi^2 / 4E$ is the minimum pion energy
required to produce a photon of energy $E$, and similarly $E_{\rm
p,th}(E_\pi)$ is the minimum energy of protons for pion production.  The
density of ICM, $n_{\rm H}(r)$, is very well measured by the X-ray
observations of bremsstrahlung radiation from thermal electrons, and the
cross section of the proton--proton collision for pion production, ${\rm
d} \sigma_{\rm pp} / {\rm d}E_\pi$, can be calibrated using laboratory
data. The distribution function of cosmic-ray protons $n_{\rm p}(E_{\rm
p}, r)$ depends on the injection power, spectrum, and spatial
distribution of cosmic rays.  By specifying these ingredients, we can
predict the $\gamma$-ray flux from a cluster.

In practice, we use a fitting formula as well as cluster parameters
given in \citet{Pfrommer2004}; for the former, we briefly summarize it
in Appendix~\ref{app:pion}.
In addition, one should also note that electrons and positrons produced
by $\pi^\pm$ decays can scatter cosmic microwave background (CMB)
photons up to $\gamma$-ray energies.
For a while, we neglect this secondary process, but revisit it in
Section~\ref{sec:IC} and show that it is in fact negligible under most
of realistic situations.

\subsection{Cosmic-ray power and spectrum}
\label{sub:Cosmic-ray power and spectrum}

The cosmic-ray pressure $P_{\rm p}$ and energy density $\rho_{\rm p}$,
which are the quantities that we want to constrain, are directly
related to the injection power of the cosmic rays.  The cosmic-ray
spectrum is measured to be a power law with the index of $\alpha_{\rm
p} = 2.7$ in our Galaxy, but in the clusters it is perhaps harder,
since they can confine cosmic rays for cosmological times
\citep*{Volk1996,Ensslin1997, Berezinsky1997}.  We thus adopt harder
spectrum with $\alpha_{\rm p} = 2.1$ and 2.4, but also use
$\alpha_{\rm p} = 2.7$ as a limiting case.

It is also possible that the injection of the cosmic rays and thus
their energy density $\rho_{\rm p}$ are intermittent.  Although it is
interesting to constrain the source property by measuring such
$\gamma$-ray variability, this is not the primary focus in the present
paper.  Instead, we concentrate on constraining energy density
$\rho_{\rm p}$ averaged over {\it GLAST} exposure time.  For the
sensitivity of {\it GLAST}, we consider the result of one-year all-sky
survey, which corresponds to $\sim$70-day exposure to each source as
the field of view is $\sim$20\% of the whole sky.  Therefore, any time
variability within this 70-day duration is smeared out.

\subsection{Radial distribution}
\label{sub:Radial distribution}

We define quantities $X_{\rm p}$ and $Y_{\rm p}$ as ratios of energy
density and pressure of cosmic rays to those of thermal gas,
respectively, i.e.,
\begin{equation}
 X_{\rm p} \equiv \frac{\rho_{\rm p}}{\rho_{\rm th}},~~
 Y_{\rm p} \equiv \frac{P_{\rm p}}{P_{\rm th}}.
\end{equation}
In general, these depend on the radius, but the concrete dependence is
totally unknown.  Various mechanisms supplying the cosmic-ray protons
have been proposed, which produce characteristic and diverse profiles
of $X_{\rm p}$ and $Y_{\rm p}$.  We thus parameterize them using a
simple power-law:
\begin{eqnarray}
 X_{\rm p}(r) & = & X_{\rm p}(R_{\rm 500})
  \left(\frac{r}{R_{\rm 500}}\right)^\beta,
  \nonumber\\
 Y_{\rm p}(r) & = & Y_{\rm p}(R_{\rm 500})
  \left(\frac{r}{R_{\rm 500}}\right)^\beta,
  \label{eq:profile}
\end{eqnarray}
where $R_{\Delta}$ (here $\Delta = 500$) is the radius at which the
enclosed spherical overdensity is $\Delta$ times the critical density
of the Universe at the cluster's redshift,\footnote{We use $R_{\Delta}
h E(z)=r_5 (T_{\rm spec}/5~{\rm keV})^{\eta / 3}$, where $r_5 = 0.792
~ h^{-1} ~ {\rm Mpc}$ and $\eta = 1.58$ for $\Delta = 500$, $r_5 =
0.351 ~ h^{-1} ~ {\rm Mpc}$ and $\eta = 1.64$ for $\Delta = 2500$, and
$E^2(z) = \Omega_{\rm m} (1+z)^3 + \Omega_{\Lambda}$ for the flat
$\Lambda$CDM cosmology \citep{Vikhlinin2006}.} where the cluster mass
$M_\Delta$ is traditionally defined with the X-ray and SZE
measurements.  We note that this approach ignores boosts in
$\gamma$-ray flux caused by clumpiness.  The constraints derived using
a smooth model hence provide a conservative upper limit on $X_{\rm p}$
and $Y_{\rm p}$.

We first focus on $X_{\rm p}$, and later discuss $Y_{\rm p}$.
The relation between $\gamma$-ray intensity and $X_{\rm p}$ is
summarized in Appendix~\ref{app:pion} and that between $X_{\rm p}$ and
$Y_{\rm p}$ is discussed in Section~\ref{sub:Cluster mass
measurements}.
We shall study the dependence of results on $\beta$, for which we adopt
1, 0, and $-0.5$.  Below, we outline several models that motivate
these values of $\beta$.

\subsubsection{Isobaric model}

The simplest model is based on the assumption
of $\beta = 0$, i.e., the energy density of cosmic rays precisely
traces that of thermal gas everywhere in the cluster.  The latter is
proportional to temperature times number density of the thermal gas,
both of which are very well measured with X-rays for various nearby
clusters.  The gas density profile is nearly constant within a
characteristic core radius $r_{\rm c}$, beyond which it decreases as a
power law, while temperature profile is almost constant.  The core
radius and outer profile are $r_{\rm c} = 300$ kpc, $r^{-2.3}$ (Coma),
$r_{\rm c} = 200$ kpc, $r^{-1.7}$ (Perseus), and $r_{\rm c} = 20$ kpc,
$r^{-1.4}$ (Virgo) \citep[see Table~1 of][for a more comprehensive
list]{Pfrommer2004}.  The latter two clusters have an even smaller `cool
core', but this structure gives only a minor effect on the $\gamma$-ray
flux.

\subsubsection{Large-scale structure (LSS) shocks}

The formation of galaxy
clusters is due to merging or accretion of smaller objects.  When
this occurs, the shock waves are generated at the outskirts of the
clusters, somewhere around $\sim$3 Mpc from the center, where protons
and electrons are accelerated to relativistic energies
\citep[e.g.,][]{Loeb2000,Miniati2002,Keshet2003,Gabici2003}.  Unlike
electrons that immediately lose energies through synchrotron radiation
and inverse-Compton (IC) scattering off CMB photons, protons are hard to
lose energies, and they are
transported efficiently into the cluster center following the motion
of ICM gas \citep{Miniati2001}.  In order to predict the eventual
profile of the cosmic-ray energy density, one needs to resort to
numerical simulations.  The recent radiative simulations by
\citet{Pfrommer2007} show somewhat jagged shape for the $X_{\rm p}(r)$
profile, which implies large clumping factor.  Here, we model its
global structure with a smooth profile with $\beta = -0.5$, ignoring
the effects of clumpiness.  On the other hand, they also performed
nonradiative simulations which rather imply $\beta = 1$ profile.
Although the latter may not be realistic, the effects of cooling and
heating in clusters are also somewhat uncertain.  Thus, we still adopt
this model, treating it as an extreme case.

\subsubsection{Central point source}

A central powerful source such as active galactic nuclei or cD galaxy
might be an efficient supplier of the cosmic rays, which diffuse out
from the central region after injection.  The profile of cosmic-ray
energy density is $r^{-1}$, but truncated at a radius that is far
smaller than $R_{500}$ for relevant energies
\citep{Berezinsky1997,Colafrancesco1998}.  The actual $\gamma$-ray
detection might therefore cause significant overestimate of the
cosmic-ray pressure; we address this issue in Section~\ref{sub:Direct
constraint from large radii}.

Numerical simulations of jets from active galactic nuclei suggest that
temporal intermittency and spatial structure might be complicated
\citep[e.g.,][]{O'Neill2005}.  Neither of these, however, affect our
results that depend on global and time-averaged properties.

\section{Cosmic-ray energy density in nearby galaxy clusters}
\label{sec:Results}

\subsection{Constraints from entire region of clusters}
\label{sub:Nearby individual clusters}

\begin{figure}
\begin{center}
\includegraphics[width=8.4cm]{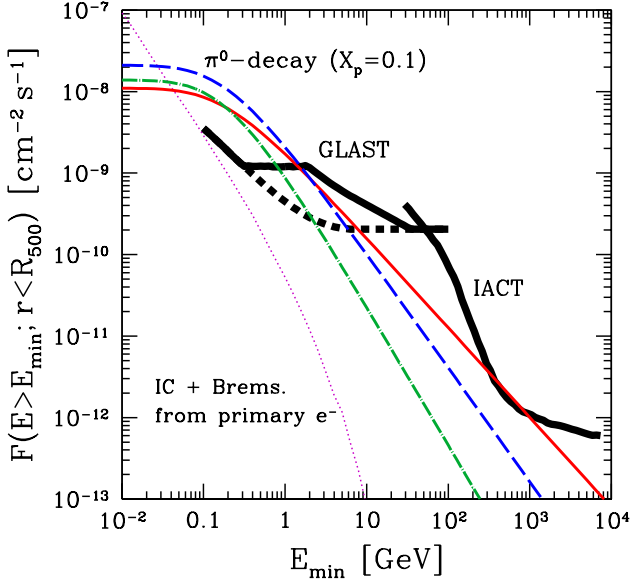}
\caption{Flux of $\gamma$-ray emission from the region within $R_{500} =
 2.1$ Mpc of the Coma cluster, for the isobaric model with $X_{\rm p} =
 0.1$ (labeled as `$\pi^0$-decay').  The spectral index of the
 cosmic-ray protons is $\alpha_{\rm p} = 2.1$ (solid), 2.4 (dashed), and
 2.7 (dot-dashed).  The sensitivity curves of {\it GLAST} and IACTs are
 for a source extended by $\theta_{500} = 1\fdg2$ (corresponding to
 $R_{500}$), while the point-source sensitivity of {\it GLAST} is also
 shown as a short dashed curve.  Flux due to IC scattering
 and nonthermal bremsstrahlung is also shown \citep[dotted;
 from][]{Reimer2004}.}
\label{fig:spectrum}
\end{center}
\end{figure}

\begin{table*}
\caption{Sensitivity to $X_{\rm p}(R_{500})$ of {\it GLAST}
 ($E_{\rm min} = 100$ MeV) for various values of spectral index of
 cosmic rays $\alpha_{\rm p}$, and isobaric and LSS shock models for
 radial distribution or $\beta$.  The limits on $X_{\rm p}$ are set by
 the $\gamma$-ray flux from a region within whichever of the larger
 between the point spread function $\delta \theta_{\rm PSF}(E_{\rm
 min}) \approx 3\degr$ and the source extension $\theta_{500}$.}
\label{table:sensitivity}
\begin{tabular}{@{}lclllllllllll}\hline
 & & \multicolumn{3}{c}{$X_{\rm p,lim}(R_{500})$ for
 $\beta = 1$} & & \multicolumn{3}{c}{$X_{\rm p,lim}(R_{500})$ for $\beta
 = 0$} & & \multicolumn{3}{c}{$X_{\rm p,lim}(R_{500})$ for $\beta =
 -0.5$}\\
 \cline{3-5}\cline{7-9}\cline{11-13}
 Cluster & $\theta_{500}$ & $\alpha_{\rm p} = 2.1$ &  $\alpha_{\rm p} =
 2.4$ & $\alpha_{\rm p} = 2.7$ & & $\alpha_{\rm p} = 2.1$ &
 $\alpha_{\rm p} = 2.4$ & $\alpha_{\rm p} = 2.7$ & & $\alpha_{\rm p} =
 2.1$ & $\alpha_{\rm p} = 2.4$ & $\alpha_{\rm p} = 2.7$ \\ \hline
 Coma & 1$\fdg$2 & 0.11 & 0.063 & 0.10 & & 0.040 & 0.022 & 0.035 &
 & 0.018 & 0.0098 & 0.016 \\
 Perseus & 1$\fdg$5 & 0.024 & 0.013 & 0.022 & & 0.012 & 0.0068 & 0.011
 & & 0.0050 & 0.0027 & 0.0044 \\
 Virgo & 4$\fdg$6 & 0.076 & 0.042 & 0.067 & & 0.041 & 0.022 & 0.036 &
 & 0.016 & 0.0088 & 0.014 \\
 Ophiuchus & 1$\fdg$3 & 0.088 & 0.048 & 0.078 & & 0.020 & 0.011 &
 0.018 & & 0.0064 & 0.0035 & 0.0056 \\
 Abell 2319 & 0$\fdg$6 & 0.048 & 0.027 & 0.043 & & 0.057 & 0.031 &
 0.050 & & 0.032 & 0.018 & 0.029 \\
 \hline
\end{tabular}
\end{table*}

We first discuss the case of the Coma cluster, focusing on the region
within $R_{500} = 2.1$ Mpc and assuming the isobaric distribution of the
cosmic-ray energy density ($\beta = 0$).  Fig.~\ref{fig:spectrum} shows
the integrated $\gamma$-ray flux with photon energies above $E_{\rm
min}$, $F(>E_{\rm min})$, for $X_{\rm p} = 0.1$.  This flux is to be
compared with the sensitivities of {\it GLAST} and IACTs, for which one
has to take the source extension into account.  Indeed, the radial
extension of the Coma cluster $R_{500}$ corresponds to $\theta_{500} =
1\fdg2$, which at high energies exceeds the size of the point spread
function (PSF), $\delta \theta_{\rm PSF} (E)$.  We obtain the flux
sensitivity for an extended source from that for a point source by
multiplying a factor of $\max[1, \theta_{500} / \delta \theta_{\rm PSF}
(E_{\rm min})]$, if the sensitivity is limited by backgrounds.  On the
other hand, if the expected background count from the cluster region is
smaller than one, which is the case for {\it GLAST} above $\sim$30 GeV,
the sensitivities for a point source and an extended source are
identical.  The region $\sim$2--30 GeV is where the expected background
count is smaller than one from the PSF area but larger than one from the
entire cluster.
We assume that IACTs are limited by background over the entire energy
region, and we multiply the point source sensitivity by $ \theta_{500} /
\delta \theta_{\rm PSF} $ with $\delta \theta_{\rm PSF} =
0\fdg1$; this is consistent with \citet{Aharonian1997} for relevant
energy regime.
A more detailed derivation of this sensitivity is given in
Appendix~\ref{app:sensitivity}.

We also show flux of IC scattering and bremsstrahlung radiations from
electrons primarily accelerated in the shocks \citep{Reimer2004}.  The
authors suggested that these electron components would always be below
the {\it GLAST} and IACT sensitivities, based on constraints from radio,
extreme-ultraviolet (EUV), and hard X-ray observations.  If this is the
case, the $\gamma$-ray detection would imply existence of cosmic-ray
protons, and be used to constrain the pressure from this component
\citep*[see also,][]{Ensslin1999,Atoyan2001}.
We give more detailed discussions about IC mechanisms in
Section~\ref{sec:IC}.

Fig.~\ref{fig:spectrum} shows that $\gamma$-rays from $\pi^0$ decays are
detectable for $X_{\rm p}=0.1$.  In particular, the models with
different values of $\alpha_{\rm p}$ predict similar amount of
$\gamma$-ray fluxes for low-energy thresholds ($E_{\rm min} < 1$ GeV);
{\it GLAST} measurements can therefore provide constraints on
$X_{\rm p}$, almost independent of $\alpha_{\rm p}$.  Non-detection with
{\it GLAST} from these nearby clusters is also very interesting as it
provides very tight upper limit to the cosmic-ray energy density in
clusters.  The fluxes above $\sim$1 TeV, on the other hand, depends very
sensitively on $\alpha_{\rm p}$;  IACTs will thus constrain the spectral
index.

In Table~\ref{table:sensitivity}, we summarize the sensitivity to
$X_{\rm p}(R_{500})$ for {\it GLAST} in the case of $E_{\rm min} =
100$ MeV, for several values of $\alpha_{\rm p}$ and different models
of radial distribution of cosmic-ray energy density.  We also
performed the same analysis for other nearby rich clusters (Perseus,
Virgo, Ophiuchus, and Abell 2319), and report their results as well.
This indeed confirms that the {\it GLAST} constraints on $X_{\rm p}$
depend only weakly on the assumed spectral index.\footnote{Note that
the sensitivity peaks at $\alpha_{\rm p} = 2.5$.  This is because for
even larger $\alpha_{\rm p}$, the contribution from low-momentum protons
to the energy density becomes more significant, while they do not
produce $\gamma$-rays.}  The constraints improve for smaller values of
$\beta$.  For $\beta \leq 0$, the {\it GLAST} non-observation can place
tight upper limits on the cosmic-ray energy density at a few per cent
level.  Even in the case of nonradiative LSS shock model ($\beta = 1$)
the constraint is still as good as $\sim$10\% for the Coma.  This is a
dramatic improvement from the EGRET bounds \citep[see, e.g., Table~3
of][]{Pfrommer2004}, by more than an order of magnitude.

On the other hand, the IACT constraints on $X_{\rm p}$ (with $E_{\rm
min} = 1$ TeV) for the Coma cluster and $\beta = 0$ profile are 0.37,
2.3, and 42 for $\alpha_{\rm p} = 2.1$, 2.4, and 2.7, respectively.
Thus, IACTs will therefore provide constraints on the spectral index,
which is directly related to astrophysical mechanisms of particle
acceleration.
A similar trend can be found in Table~6 of \citet{Pfrommer2004};
however, the authors applied point-source flux limit to the (extended)
clusters and obtained much more stringent sensitivities than ours.

\subsection{Direct constraint from large radii}
\label{sub:Direct constraint from large radii}

So far, we treated all clusters but Virgo as point sources.  Although
we showed that the dependence on the assumed radial profile was
reasonably weak, a more general approach would be to use the resolved
image.  This is particularly useful, if the radial profile cannot be
simply parameterized (see Section~\ref{sub:Radial distribution}).
Because we are interested in the cosmic-ray pressure at $R_{500}$ and
the $\gamma$-ray yields would rapidly drop with radius, we here
consider constraints in a projected radial shell between
$\theta_{2500}$ and $\theta_{500}$.  We mainly focus on the Perseus,
and assume $\alpha_{\rm p} = 2.1$; in this case, $\theta_{2500} =
0\fdg65$.  In order to resolve the inner region, we consider the
energy threshold of 0.6 GeV, above which the {\it GLAST} resolution
becomes smaller than $\theta_{2500}$.  The {\it GLAST} flux limits for
the outer region and for $E > 0.6$ GeV correspond to the following
limits on the fractional energy density: $X_{\rm p,lim} (R_{500}) =
0.099$, 0.089, and 0.080, for $\beta = 1$, 0, and $-0.5$,
respectively, which are still reasonably small.  In addition, these
are much less sensitive to the assumed profile, thus applicable to
more general cases including the central source model.  The similar
procedure predicts sensitivities for other clusters: $X_{\rm
p}(R_{500}) = 0.42$ (Coma), 0.14 (Virgo), 0.41 (Ophiuchus), and 0.55
(Abell 2319), in the case of $\beta = 0$ and $\alpha_{\rm p} = 2.1$.
Although it is limited to nearby clusters, such analysis provides an
important handle on the radial distribution of cosmic-ray ions in
clusters.

\section{Evolution of cosmic-ray energy density}
\label{sec:Evolution of cosmic-ray energy density}

While we could obtain stringent constraints for individual nearby
clusters, these rapidly get weaker for more distant clusters.  In this
case, however, one can stack many clusters to overcome the loss of
signals from each cluster.  \citet{Reimer2003} took this approach for
the EGRET analysis, and obtained an improved upper limit to the average
flux of 50 nearby clusters.
We argue that the flux is not a physical quantity because it depends on
distance and therefore distribution of sources.
We should instead convert this improved flux limit to constraint on more
physical quantities such as $\gamma$-ray luminosity.
Here, we examine the {\it GLAST} constraints on $X_{\rm p}(R_{500})$
obtained by stacking clusters from the whole sky and in several
redshift intervals.  As we consider rather distant clusters, they are
all treated as point sources.

\subsection{Stacking \boldmath{$\gamma$}-ray signals from galaxy clusters}
\label{sub:stacking}

\subsubsection{Formulation and models}

\begin{figure}
\begin{center}
\includegraphics[width=8.4cm]{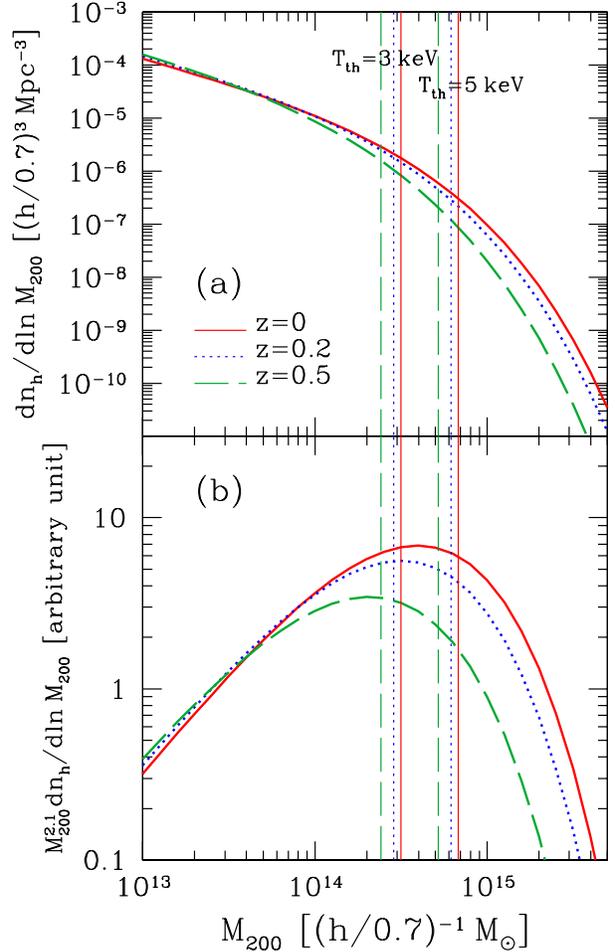}
\caption{(a) Cluster mass function as a function of $M_{200}$
 at several redshifts.  Threshold mass $M_{\rm th}$ corresponding to
 $T_{\rm th} = 3$, 5 keV is shown as vertical lines.  (b)
 Cluster mass function $\mathrm dn_{\rm h} / \mathrm d\ln M_{200}$
 multiplied by $M_{200}^{2.1} (\propto F_{\rm X_{\rm p}})$, in arbitrary
 units.  Line types are the same as in (a).}
\label{fig:massfunction}
\end{center}
\end{figure}

\begin{table*}
\caption{{\it GLAST} sensitivities to $X_{\rm p}(R_{500})$ and $Y_{\rm
 p}(R_{500})$ by stacking $N_{\rm cl}$ clusters above threshold
 temperature $T_{\rm th}$ at given redshift ranges, for $\alpha_{\rm
 p} = 2.1$, $\beta = 0$, and $E_{\rm min} = 1$ GeV.}
\label{table:stacking}
\begin{tabular}{@{}crcccrcc} \hline
 & \multicolumn{3}{c}{$T_{\rm th} = 3$ keV} & &
 \multicolumn{3}{c}{$T_{\rm th} = 5$ keV}
 \\
 \cline{2-4}\cline{6-8}
 $z$ & $N_{\rm cl}$ & $X_{\rm p,lim}$ & $Y_{\rm p,lim}$ & & $N_{\rm
 cl}$ & $X_{\rm p,lim}$ & $Y_{\rm p, lim}$
 \\
 \hline
 0.05--0.10 & 200 & 0.11 & 0.06 & & 30 & 0.09 & 0.05 \\
 0.10--0.15 & 530 & 0.21 & 0.11 & & 60 & 0.16 & 0.09 \\
 0.15--0.25 & 2500 & 0.29 & 0.16 & & 290 & 0.23 & 0.13 \\
 0.25--0.40 & 7900 & 0.57 & 0.31 & & 870 & 0.46 & 0.25 \\
 0.40--0.60 & 17000 & 1.3 & 0.72 & & 1700 & 1.1 & 0.60 \\
\hline
\end{tabular}
\end{table*}

The number of clusters with $M > M_{\rm th}$ between redshifts $z_1$ and
$z_2$ is given by
\begin{equation}
 N_{\rm cl} = \int_{z_1}^{z_2} \mathrm dz \frac{\mathrm dV}{\mathrm dz}
  \int_{M_{\rm th}}^{\infty} \mathrm dM
  \frac{\mathrm dn_{\rm h}}{\mathrm dM}(M,z),
  \label{eq:cluster number}
\end{equation}
where ${\rm d}V$ is the comoving volume element, $\mathrm dn_{\rm
h}/\mathrm dM$ is the halo mass function (comoving number density of
dark matter halos per unit mass range); the former can be computed given
cosmological parameters, and for the latter we use the following
parameterization:
\begin{equation}
\frac{\mathrm dn_{\rm h}}{\mathrm dM_{\rm 180m}}
 = A_{\rm J} \frac{\Omega_{\rm m}\rho_{\rm c}}{M_{\rm 180m}}
 \exp\left[-|\ln\sigma^{-1}+B_{\rm J}|^{\epsilon_{\rm J}}\right]
 \frac{\mathrm d\ln\sigma^{-1}}{\mathrm dM_{\rm 180m}},
\end{equation}
where $\rho_{\rm c}$ is the critical density of the present Universe,
$\sigma (M_{\rm 180m}, z)$ is a standard deviation for distribution
function of linear over density, $A_{\rm J} = 0.315$, $B_{\rm J} =
0.61$, and $\epsilon_{\rm J} = 3.8$ \citep{Jenkins2001}.
Here we note that $M_{\rm 180m}$ is defined as an enclosed mass within a
given radius, in which the average density is $180 \Omega_{\rm m}
\rho_{\rm c} (1+z)^3$.

We give the threshold mass $M_{\rm th}(z)$ in terms of threshold
temperature $T_{\rm th}$ based on the observed mass--temperature
relation: $M_{200} = 10^{15} h^{-1} M_{\sun} (T/8.2 ~
\mathrm{keV})^{3/2} E(z)^{-1}$ \citep{Voit2005}.  This is because the
efficiency of large-scale SZE cluster surveys relies mainly on cluster
temperature {\it regardless of} cluster redshifts.  Note that this
relation is between temperature and mass $M_{200}$, which is within a
radius $R_{200}$.  Here we use the prescription of \citet{Hu2003} for
the conversion of different mass definitions, $M_{200}$ and $M_{\rm
180m}$ with assumed concentration parameter $c_{\rm v} = 3$.  For the
threshold temperature, we adopt $T_{\rm th} = 3$ and
5 keV.  Fig.~\ref{fig:massfunction}(a) shows the mass function as well
as threshold mass corresponding to $T_{\rm th}$, at various redshifts.
In Table~\ref{table:stacking}, we list values of $N_{\rm cl}$ after
integrating equation~(\ref{eq:cluster number}), for several redshift
ranges and different $T_{\rm th}$.

The average flux of $\gamma$-rays from these clusters is
\begin{equation}
 \overline F_{{\rm st},X_{\rm p}} = \frac{1}{N_{\rm cl}}
  \int_{z_1}^{z_2} {\rm d}z \frac{{\rm d}V}{{\rm d}z}
  \int_{M_{\rm th}}^{\infty} \mathrm dM
  \frac{\mathrm dn_{\rm h}}{\mathrm dM}(M,z)
  F_{X_{\rm p}}(M,z),
  \label{eq:average flux}
\end{equation}
where $F_{X_{\rm p}}(M,z)$ is the $\gamma$-ray flux from a cluster of
mass $M$ at redshift $z$, given $X_{\rm p}$.
The flux from each cluster above $E_{\rm min}$ is written as
\begin{equation}
 F_{X_{\rm p}}(M,z) = \frac{1+z}{4\pi d_{\rm L}^2}
  \int \mathrm dV_{\rm cl}
  \int_{(1+z)E_{\rm min}}^\infty \mathrm dE\
  q_\gamma (E,r|M),
\end{equation}
where $d_{\rm L}$ is the luminosity distance, $\mathrm dV_{\rm cl}$
represents the cluster volume integral, and $q_\gamma$ is the volume
emissivity given by equation~(\ref{eq:emissivity}) or
(\ref{eq:emissivity 2}).

\begin{figure}
\begin{center}
\includegraphics[width=8.4cm]{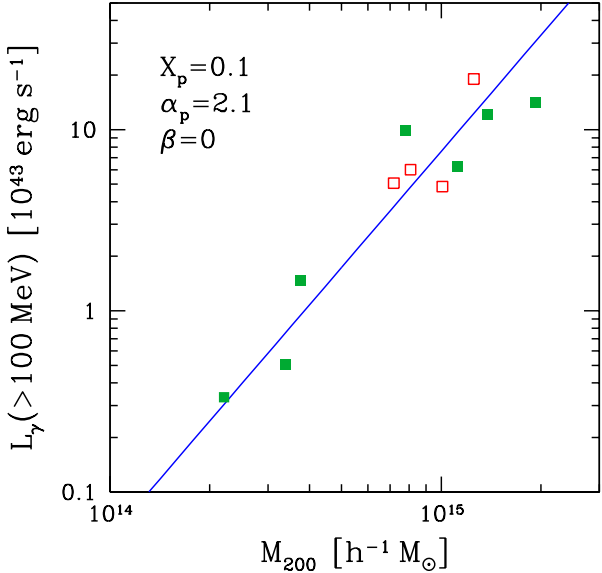}
\caption{Relation between $\gamma$-ray luminosity (above 100 MeV) and
 cluster mass $M_{200}$, for several nearby clusters and for the
 parameters $X_{\rm p} = 0.1$, $\alpha_{\rm p} = 2.1$, and $\beta =
 0$. Filled (open) points are for cooling flow (non-cooling flow)
 clusters. The solid line is the $L_\gamma \propto M_{200}^{2.1}$
 profile that fits the data quite well.}
\label{fig:cluster_lum}
\end{center}
\end{figure}

We then quantify the mass dependence of this flux $F_{X_{\rm
p}}(M,z)$.  In the case of the isobaric model ($\beta = 0$) with a
fixed $X_{\rm p}$, the $\gamma$-ray luminosity scales as ICM number
density times energy density, i.e., $L_\gamma \propto X_{\rm p} n_{\rm
H} \rho_{\rm th} \propto X_{\rm p} n_{\rm H}^2 T$.  On the other hand,
luminosity of X-rays due to the thermal bremsstrahlung process scales
as $L_{\rm X} \propto n_{\rm H}^2 T^{1/2}$.  Therefore, there is a
relation between $\gamma$-ray and X-ray luminosities as follows:
$L_\gamma / L_{\rm X} \propto X_{\rm p} T^{1/2}$.  In addition, there
are empirical relations between X-ray luminosity and cluster mass,
$L_{\rm X} \propto M_{200}^{1.8}$, and also between gas temperature
and mass, $T \propto M_{200}^{2/3} E^{2/3}(z)$ \citep{Voit2005}.  Thus,
combining these three and assuming that $X_{\rm p}$ is independent of
mass, we obtain a scaling relation $L_\gamma \propto X_{\rm p}
M_{200}^{2.1} E^{1/3}(z)$.  In
Fig.~\ref{fig:cluster_lum}, we show predicted $\gamma$-ray luminosity
as a function of cluster mass (inferred from temperature), for several
well-measured nearby clusters (taken from tables in
\citealt{Pfrommer2004}) with the parameters $X_{\rm p} = 0.1$,
$\alpha_{\rm p} = 2.1$, and $\beta = 0$.  The $L_\gamma$--$M_{200}$
relation can indeed be well fitted with $L_\gamma (> 100 ~ {\rm MeV})
= 7.6 \times 10^{44} X_{\rm p} (M_{200} / 10^{15} h^{-1}
M_{\sun})^{2.1}$ erg s$^{-1}$ for clusters at $z \approx 0$, shown as a
solid line in Fig.~\ref{fig:cluster_lum}.  When we compute the
$\gamma$-ray flux $F_{X_{\rm p}}(M|z)$ (or equivalently luminosity)
from clusters with a given mass $M$, we adopt this mass--luminosity
relation as a model for average cluster, and scale as $L_\gamma \propto
E^{1/3}(z)$ for high-redshift clusters.

Fig.~\ref{fig:massfunction}(b) shows the mass function weighed by the
mass dependence of the flux (in arbitrary unit).  This quantity
represents which mass scale dominates the average flux at each
redshift.  From this figure, one can see that clusters with $M_{200}
\sim 3 \times 10^{14} M_{\sun}$ most effectively radiates
$\gamma$-rays in the low-redshift Universe, but the distribution is
rather broad for $\sim$10$^{14}$--10$^{15}M_{\sun}$.  If we adopt
$T_{\rm th} = 5$ keV, then the clusters around the threshold mass are
the more dominant contributors to the average flux.

\subsubsection{\textit{GLAST} constraints on $X_{\rm p}$}

The average flux of the stacked clusters (equation~\ref{eq:average
flux}) is then compared with the corresponding {\it GLAST}
sensitivity,
\begin{equation}
 F_{\rm st,lim} = \frac{F_{\rm lim}}{\sqrt{N_{\rm cl}}}.
  \label{eq:GLAST sensitivity}
\end{equation}
where $F_{\rm lim}$ is the sensitivity to each cluster given as the
thick dashed line in Fig.~\ref{fig:spectrum} (for a point-like
source).  To derive constraints on $X_{\rm p}$ from the stacked image,
we solve $\overline F_{{\rm st},X_{\rm p}} = F_{\rm st,lim}$ for
$X_{\rm p}$.  Throughout the following discussion, we adopt $\beta = 0$,
$\alpha_{\rm p} = 2.1$ and $E_{\rm min} = 1$ GeV, around which the
$\gamma$-ray yields are maximized compared with the point-source
sensitivity (Fig.~\ref{fig:spectrum}).  In addition, the pixel number
with this threshold ($4 \pi$ divided by PSF area; $6\times 10^{4}$) is
large enough to minimize the positional coincidence of multiple
clusters (compare with $N_{\rm cl}$'s in Table~\ref{table:stacking}).

We summarize the results in Table~\ref{table:stacking}.  We find that
the limits are as strong as $X_{\rm p}\lesssim 0.16$ (0.23) for $0.1 < z
< 0.15$ ($0.15 < z < 0.25$).  The sensitivities improve for larger
$T_{\rm th}$, because the smaller cluster number is compensated by the
strong mass dependence of the flux.  The constraints on $X_{\rm p}$
degrades rapidly with redshift.  Table~\ref{table:stacking} also shows
{\it GLAST} sensitivities for $Y_{\rm p}$, which is almost twice as
stringent as those for $X_{\rm p}$ in the case of $\alpha_{\rm p} =
2.1$.  We discuss implications of this result for $Y_{\rm p}$ in
Section~\ref{sub:Cluster mass measurements} in details.

The current X-ray catalog covers clusters at $z \lesssim 0.2$ for
$T_{\rm th} = 5$ keV \citep{Boehringer2001}.  The {\it GLAST} data
could thus immediately be compared with this low-redshift catalog.  At
higher redshifts, the South Pole Telescope would find many clusters
with $T \gtrsim 3$ keV using SZE; but since it covers $\sim$10\% of
the whole sky, the limits would become $\sim$3 times weaker than those
in Table~\ref{table:stacking}.  The {\it Planck} satellite, on the
other hand, would yield all-sky SZE catalog of very massive clusters;
we find that the limits for $T_{\rm th} = 8$~keV clusters are nearly
identical to those for $T_{\rm th} = 5$~keV systems.  

In addition to probing its redshift evolution, the stacking approach
is also useful for studying cosmic-ray component in nearby low-mass
clusters, and the dependence of $X_{\rm p}$ on cluster mass.
Although individual clusters are not bright enough,
cluster mass function predicts that there are a number of such
low-mass clusters, which should help improve the {\it GLAST}
sensitivity.

\subsection{Extragalactic \boldmath{$\gamma$}-ray background}

Another avenue to constrain the universal average of $X_{\rm p}$ is to
use the extragalactic $\gamma$-ray background \citep{Sreekumar1998},
because galaxy clusters would contribute to this background intensity to
a certain extent.
Their contribution is quantified as
\begin{equation}
 I_\gamma = \int_0^\infty \mathrm dz
  \frac{\mathrm d^2V}{\mathrm dz \mathrm d\Omega}
  \int_{M_{\rm th}}^{\infty} \mathrm dM
  \frac{\mathrm dn_{\rm h}}{\mathrm dM}(M,z)F_{X_{\rm p}}(M,z),
  \label{eq:EGB}
\end{equation}
which is quite similar to equation~(\ref{eq:average flux}).
Adopting the same models for $\mathrm dn_{\rm h} / \mathrm dM$ and
$F_{X_{\rm p}}$ as in Section~\ref{sub:stacking}, and
using $\alpha_{\rm p} = 2.1$, $\beta = 0$, and $E_{\rm min} = 100$ MeV,
we obtain
\begin{equation}
 I_\gamma (>100~{\rm MeV}) = 4\times 10^{-7} X_{\rm p}
  ~\mathrm{cm^{-2}~s^{-1}~sr^{-1}}.
\end{equation}
Even with $X_{\rm p} = 1$, this is much smaller than the measurement
by EGRET: $10^{-5}$ cm$^{-2}$ s$^{-1}$ sr$^{-1}$
\citep{Sreekumar1998}.
This indicates that cosmic-ray processes in galaxy clusters are very
unlikely to contribute to the $\gamma$-ray background flux
significantly, especially because it requires a very large value for
$X_{\rm p}$, which is already excluded by EGRET for some of nearby
clusters.
This result is consistent with the previous studies such as
\citet{Colafrancesco1998}.
Hence, we conclude that the stacking method using resolved clusters
introduced in Section~\ref{sub:stacking} would provide much more
stringent constraint on $X_{\rm p}$ than the approach using
extragalactic $\gamma$-ray background.

However, we here mention a few possibilities that may render this
approach more viable in the near future.
Soon after launch, {\it GLAST} should start resolving many point sources
(mainly blazars) that are now contributing to the background flux.
Furthermore, using angular power spectrum of the $\gamma$-ray background
map might enable to disentangle the origin \citep{Ando2006,Ando2007b}.
In addition, there is a claim that the measured $\gamma$-ray background
flux is dominated by the Galactic foreground even at high latitude, and
that there is no certain measurement of truly extragalactic component
\citep*{Keshet2004}.
In any of the cases above, the contribution from galaxy clusters might
be found to be significantly smaller than the current observed flux,
which would be useful to constrain $X_{\rm p}$ at higher redshifts.

\section{X-ray and SZE cluster mass estimates}
\label{sub:Cluster mass measurements}

\begin{figure}
\begin{center}
\includegraphics[width=8.4cm]{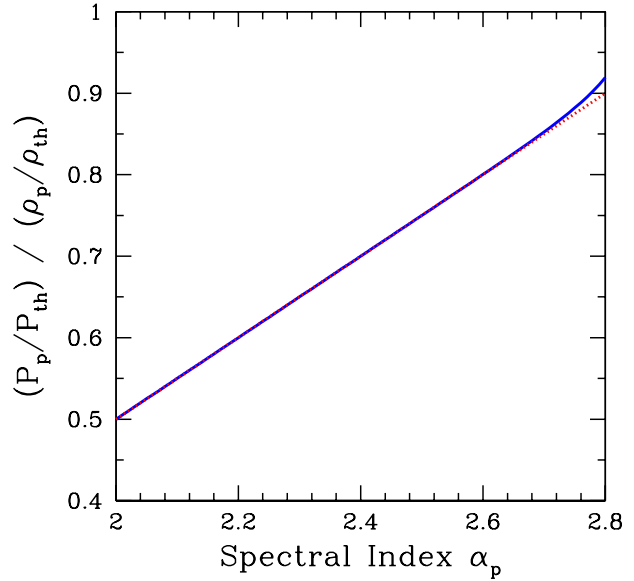}
\caption{Relation between ratios of pressure ($Y_{\rm p} = P_{\rm p} /
  P_{\rm th}$) and energy density ($X_{\rm p} = \rho_{\rm p} /
  \rho_{\rm th}$) plotted as a function of spectral index $\alpha_{\rm
  p}$ of cosmic rays (solid line).  Dotted line is the linear fit
  $Y_{\rm p} / X_{\rm p} = 0.5 (\alpha_p - 1)$.}
\label{fig:pres_corr}
\end{center}
\end{figure}

Future $\gamma$-ray observations of galaxy clusters will have the
potential to place tight constraints on the nonthermal pressure
provided by cosmic rays.  These forthcoming $\gamma$-ray constraints
will, in turn, provide important handle on systematic uncertainties in
the X-ray and SZE cluster mass estimates based on the hydrostatic
equilibrium of the ICM.  The hydrostatic mass profile of a
spherically-symmetric cluster is given by
\begin{equation}
M(<r) = \frac{-r^2}{G\rho_{\rm g}} \left( \frac{\mathrm dP_{\rm
th}}{\mathrm dr}+\frac{\mathrm dP_{\rm nt}}{\mathrm dr} \right),
\label{eq:HSE}
\end{equation}
where $M(<r)$ is the mass enclosed within radius $r$, $\rho_{\rm g}$
is the gas density, and $P_{\rm th}$ and $P_{\rm nt}$ are the thermal
and the nonthermal contributions to the pressure.
The thermal gas, measured directly with current X-ray and SZE
observations, provides a significant
fraction of the total pressure support.  The contribution of the
nonthermal pressure, on the other hand, is customarily assumed to be
small ($\lesssim 10\%$) outside of a cluster core \citep*[see
e.g.,][]{Nagai2007b}, and it is often ignored in the hydrostatic mass
estimates based on X-ray and SZE data.  The cosmic-ray pressure, if
present, is a potential source of systematic bias in the hydrostatic
mass estimates of clusters \citep[e.g.,][and references
therein]{Ensslin1997,Rasia2006,Nagai2007a}.

In equation~(\ref{eq:HSE}), a directly relevant quantity is pressure
gradient rather than energy density $X_{\rm p}$ that we mainly discussed
until this point.
Currently, it is not possible to infer both pressure and its radial
profile, and here, we simply assume that the cosmic-ray pressure profile
is the same as that of thermal pressure.
In this case, one needs to relate $X_{\rm p}$ to $Y_{\rm p}$.
If the cosmic rays are dominated by relativistic component, then
equation of state would be $P_{\rm p} = \rho_{\rm p} / 3$.  On the other
hand, for nonrelativistic thermal gas, it is $P_{\rm th} = 2 \rho_{\rm
th} / 3$.  Thus, we expect $P_{\rm p} / P_{\rm th} = (1/2) (\rho_{\rm p}
/ \rho_{\rm th}) = X_{\rm p} / 2$.  More precisely, we can obtain the
equation of state for cosmic-ray protons by numerically integrating
the following expressions:
\begin{equation}
\rho_{\rm p} = \int_0^\infty \mathrm dp\
f_{\rm p}(p) \left(\sqrt{p^2+m_{\rm p}^2}-m_{\rm p}\right),
\end{equation}
\begin{equation}
P_{\rm p} = \int_0^\infty \mathrm dp\
f_{\rm p}(p) \frac{p^2}{3\sqrt{p^2+m_{\rm p}^2}},
\end{equation}
where $f_{\rm p}(p) \propto p^{-\alpha_{\rm p}}$ is the differential
number density distribution.  In Fig.~\ref{fig:pres_corr}, we show a
correction factor between the pressure ratio $Y_{\rm p}$ and $X_{\rm
p}$, as a function of spectral index $\alpha_{\rm p}$.  This relation
is well fitted by a linear formula $Y_{\rm p} / X_{\rm p} = 0.5
(\alpha_{\rm p}-1)$ as shown as a dotted line in
Fig.~\ref{fig:pres_corr}; the deviation is only $\sim$0.3\% at
$\alpha_{\rm p} = 2.7$.  As expected, for $\alpha_{\rm p}$ close to 2,
the ratio is about 0.5.  Therefore, the expected sensitivity of {\it
GLAST} for $Y_{\rm p}$ would be stronger than that for $X_{\rm p}$
given in Table~\ref{table:sensitivity} and as explicitly shown in
Table~\ref{table:stacking}.  For $\alpha_{\rm p} = 2.1$, {\it GLAST}
sensitivities to $Y_{\rm p}$ based on the cluster stacking method are
5\%, 9\%, and 13\% at $0.05 < z < 0.10$, $0.10 < z < 0.15$, and $0.15
< z < 0.25$, respectively.  Note, however, that the conversion between
$Y_{\rm p}$ and $X_{\rm p}$ depends on $\alpha_{\rm p}$, for which
IACT measurements would be essential.

Observational constraints on $X_{\rm p} = \left< \rho_{\rm p}
\right> / \left< \rho_{\rm th} \right>$ is also sensitive to any
non-negligible small-scale structure in the ICM.  When gas clumps, it
has density higher than the local average, $\left < \rho_{\rm th}
\right >$.  If it is not resolved and masked out, the local
inhomogeneity in the ICM boosts $\gamma$-ray surface brightness by a
factor of $C_{\gamma}\equiv \left< \rho_{\rm p} \rho_{\rm th}
\right>/\left< \rho_{\rm p} \right>\left< \rho_{\rm th} \right>$ and
X-ray surface brightness by $C_{\rm X}\equiv \left< \rho_{\rm th}^2
\right>/\left< \rho_{\rm th} \right>^2$, while leaving SZE signal
(which is linearly proportional to $\rho_{\rm th}$) unaffected by
clumpiness.  A joint $\gamma$-ray+X-ray constraints on $X_{\rm p}$
based on a smooth model is generally biased by a factor
$C_{\gamma}/C_{\rm X}$, which could be greater or less than $1$
depending on the relative size of $C_{\gamma}$ and $C_{\rm
X}$.\footnote{Current X-ray observations with superb spatial
resolution and sensitivity are capable of detecting the prominent
clumps that contribute significantly to the X-ray surface brightness.
A comparison of recent X-ray and SZE measurements indicate that the
X-ray clumping factor is very close to unity ($1<C_{\rm X}\lesssim
1.1$) in practice \citep{LaRoque06}.}  A joint $\gamma$-ray+SZE
constraint on $X_{\rm p}$, on the other hand, is biased high by a
factor $C_{\gamma}$.  Recent cosmological simulations of clusters that
include cosmic-ray physics indicate jagged shape for the $X_{\rm
p}(r)$ profile, which implies a large clumping $C_{\gamma}$
\citep{Pfrommer2007}.  These simulations are potentially useful for
estimating the values of $C_{\gamma}$, which would be important for
interpretation of $X_{\rm p}$ in case of detection of cluster signals
with upcoming $\gamma$-ray experiments.  In absence of these
constraints, observational constraints on $X_{\rm p}$ should be taken
as an {\it upper limit}.

Recently, \citet{Mahdavi07} performed a comparison between masses
estimated with weak gravitational lensing and using the assumption of
hydrostatic equilibrium, and showed that the latter masses are
typically biased to be lower by 20\%.  This result might indicate
presence of the nonthermal pressure component.  Upcoming $\gamma$-ray
measurements of galaxy clusters could thus provide useful information
on the origin of this mass discrepancy.  Turbulence and magnetic
fields are also potential sources of bias in X-ray and SZE cluster
mass estimates.  Recent numerical simulations of cluster formation
indicate that sub-sonic motions of gas provide nonthermal pressure in
clusters by about $\sim$10\% even in relaxed clusters
\citep[e.g.,][and references therein]{Rasia2006,Nagai2007a}.  Most
cluster atmospheres are also magnetized with typical field strengths
of order a few $\mu$G out to Mpc radii \citep{Carilli2002,Govoni2004},
but this would only give negligible contribution to the total pressure
support.

\section{Inverse-Compton scattering from nonthermal electrons}
\label{sec:IC}

\subsection{Secondary electrons from pion decays}

Until this point, we have neglected the contribution to $\gamma$-rays
from relativistic electrons and positrons produced from decays of
charged pions. Those charged pions are produced by the proton--proton
collisions just as $\pi^0$'s that decay into $\gamma$-rays.  Thus, as
long as the cosmic-ray protons exist, there should also be relativistic
e$^\pm$ component associated with them.  GeV $\gamma$-rays would be
produced by IC scattering of CMB photons due to such a `secondary'
leptonic component.  In this subsection, we show the expected IC flux to
compare it with the flux from $\pi^0$ decays, and argue that the
former is indeed negligible, justifying our earlier treatment.

Unlike protons, leptons can cool quickly by synchrotron radiation and
IC scattering.  Energy distribution of these electrons (positrons)
after cooling is obtained as a steady-state solution of the transport
equation, which is
\begin{equation}
n_{\rm e}(E_{\rm e},r) = \frac{1}{|\dot E_{\rm e}(E_{\rm e},r)|}
\int_{E_{\rm e}}^{\infty} \mathrm dE_{\rm e}^\prime
Q_{\rm e}(E_{\rm e}^\prime,r),
\label{eq:electron distribution}
\end{equation}
where $Q_{\rm e}$ is the source function of injected electrons.
For the energy-loss rate $\dot E_{\rm e}$, the dominant interaction
would be synchrotron radiation and IC scattering of CMB photons, i.e.,
$- \dot E_{\rm e} \propto (U_{\rm B} + U_{\rm CMB}) E_{\rm e}^2$,
where $U_{\rm B}$ and $U_{\rm CMB}$ are the energy densities of
magnetic fields and CMB.
If the injection spectrum is power law, $Q_{\rm e} \propto E_{\rm
e}^{-\alpha_{\rm e}}$, then equation~(\ref{eq:electron distribution})
states that the spectrum after cooling would be $n_{\rm e} \propto
E_{\rm e}^{-\alpha_{\rm e}-1}$, steeper by one power.

Once we know the electron distribution we can unambiguously compute
the IC spectrum after scattering CMB photons.  In addition, in the
case of the secondary electrons, we can compute the source $Q_{\rm e}$
relatively well given the spectrum of cosmic-ray protons.  In
Appendix~\ref{app:IC}, we summarize fitting formula that we use, given
by \citet{Dolag2000} and \citet{Pfrommer2004}.  Looking at
equation~(\ref{eq:electron distribution}), in order to get the
electron distribution after cooling, we also need to know magnetic
field strength $B$ in the clusters that is relevant for synchrotron
cooling.  The estimates of $B$ range $\sim$0.1--10 $\mu$G
\citep*{Clarke2001,Fusco-Femiano2004,Rephaeli2006}, while the CMB
energy density corresponds to equivalent field strength of $B_{\rm
CMB} = 3.24 (1+z)^2$ $\mu$G.  Thus, unless $B$ is larger than or
comparable to $B_{\rm CMB}$ everywhere in the cluster, the synchrotron
cooling would not be significant, as the energy loss is proportional
to $B^2 + B_{\rm CMB}^2$.  We here assume $B = 0$ to obtain the
maximally allowed IC flux.

\begin{figure}
\begin{center}
\includegraphics[width=8.4cm]{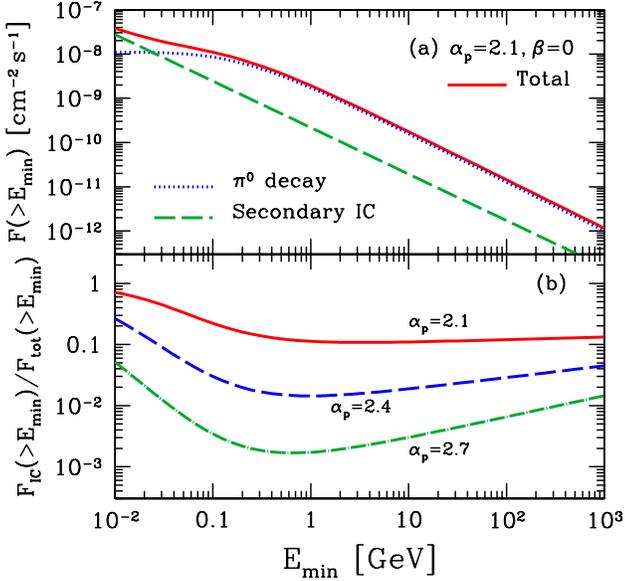}
\caption{(a) Flux of $\gamma$-rays from $\pi^0$ decays with $X_{\rm
 p}=0.1$ (dotted), IC scattering due to secondary electrons (dashed)
 as a function of minimum energy $E_{\rm min}$, for $\alpha_{\rm p} =
 2.1$, $\beta = 0$, and $B = 0$; total flux is indicated as a solid
 curve.  (b) Fractional contribution of IC scattering to the total
 $\gamma$-ray flux, $F_{\rm IC} / F_{\rm tot}$, for $\alpha = 2.1$
 (solid), 2.4 (dotted), and 2.7 (dashed).}
\label{fig:IC}
\end{center}
\end{figure}

In Fig.~\ref{fig:IC}(a), we show flux of IC $\gamma$-rays from
secondary leptons, compared with direct $\gamma$-ray flux from $\pi^0$
decays, assuming $X_{\rm p} = 0.1$, $\alpha_{\rm p} = 2.1$, and $\beta = 0$.
Fig.~\ref{fig:IC}(b) shows the fractional contribution of the IC
processes for various values of $\alpha_{\rm p}$.  These figures show
that even in the case of very week magnetic fields to reduce the
electron energy losses, the IC processes give only sub-dominant flux
in the GeV energy range relevant for {\it GLAST}.  The fractional
contribution of the IC emission to the total $\gamma$-ray flux, which is
independent of $X_{\rm p}$, is smaller than 20\% for $E_{\rm
min}=100$~MeV and $\alpha=2.1$.  For a steeper proton spectrum
($\alpha>2.1$), the fractional contribution become considerably
smaller.  Bremsstrahlung process due to the same electrons and
positrons is even more suppressed \citep{Blasi2001}.  We thus conclude
that the IC and bremsstrahlung $\gamma$-ray emission by secondary
electrons are sub-dominant for the realistic range of parameters.

\subsection{Primary electrons by shock acceleration}

Whenever the shocks are generated, both ions and electrons are
accelerated.  Thus, one expects that the IC scattering off the CMB
photons due to such primary electrons would also contribute to the
GeV--TeV $\gamma$-ray flux to a certain extent
\citep{Loeb2000,Totani2000,Waxman2000,Gabici2004}.  If this process
dominates the $\pi^0$ decays in $\gamma$-ray energy band, then the
constraints on $X_{\rm p}$ will be directly affected in case of
detection.  However, there are difficulties for this mechanism to work
efficiently in many clusters.

As electrons lose their energies via radiation much more rapidly than
protons, clusters would be bright with this mechanism during only a
limited period after injection.  For example, the radiative cooling
time scale for 10 GeV electrons is $\sim$10$^8$ years, which is much
shorter than typical cluster age.  By the same reason and also
comparing the spatial intensity distribution, it is unlikely that
synchrotron radiation from these primary electrons is responsible for
the observed radio halo emissions \citep*[e.g.,][]{Blasi2007}.

It might still be possible to overcome these difficulties if these
electrons are continuously reaccelerated in situ through the second
order Fermi mechanism
\citep*{Schlickeiser1987,Tribble1993,Brunetti2001,Petrosian2001}.  In
this case, however, the spectrum of electrons has typically a cutoff
at the Lorentz factor of $\lesssim 10^5$.  This property, while
explains spectrum of radio halo of Coma quite well
\cite[e.g.,][]{Reimer2004}, would restrict the $\gamma$-ray flux in
the GeV region due to the IC scattering and bremsstrahlung.  In
Fig.~\ref{fig:spectrum}, we show the upper bound on these components
in the case of Coma cluster as a dotted curve, taken from
\citet{Reimer2004}.

In consequence, as long as $X_{\rm p}$ is more than a few per cent, it
would be unlikely that the primary electrons, whether they are
directly injected or continuously reaccelerated, dominate the GeV
$\gamma$-ray flux, at least in a large fraction of clusters.  Even
though primary electron component dominated the detected flux, the
shape of $\gamma$-ray spectrum would be very different from
$\pi^0$-decay component especially at low energies, which could be
used as a diagnosis tool; this difference comes from the kinematics of
$\pi^0$ decays.  The {\it GLAST} energy band ranges down to $\sim$20
MeV, which is especially important characteristic for that purpose.
Moreover, observations in lower frequency bands such as radio, EUV,
and hard X-rays, are also important, because these emissions are
understood as synchrotron radiation (for radio) and IC scattering (for
EUV and hard X-rays) from nonthermal electrons.

\subsection{Secondary leptons from ultra-high energy cosmic-ray
   interactions}

If protons are accelerated up to ultra-high energies such as $\gtrsim
10^{18}$ eV in galaxy clusters, which may be plausible, these protons
are able to produce e$^\pm$ pairs through the Bethe-Heitler process
with CMB photons: ${\rm p} \gamma_{\rm CMB} \to {\rm p} {\rm e}^+ {\rm
e}^-$.  These high-energy e$^\pm$ pairs then IC scatter the CMB
photons, producing GeV--TeV $\gamma$-rays
\citep*{Aharonian2002,Rordorf2004,Inoue2005}.  In this case, the IC
photons might dominate the $\pi^0$-decay $\gamma$-rays by many orders.

However, this mechanism is extremely uncertain, depending heavily on
the maximal acceleration energy of the protons.  This is especially
because the threshold energy of the Bethe-Heitler process is
$\sim$10$^{17}$--10$^{18}$ eV, and it is unclear whether the magnetic
fields are strong enough to confine these ultra-high energy protons
for cluster ages.  Even if the detected $\gamma$-rays are dominated by
this mechanism, the spectrum would be quite different from the
$\pi^0$-decay $\gamma$-rays and should be easily distinguishable
\citep[e.g.,][]{Inoue2005}.

\section{Conclusions}
\label{sec:Conclusions}

We investigated the capability of the current and future $\gamma$-ray
detectors such as {\it GLAST} and IACTs for constraining the
cosmic-ray pressure contribution to the ICM.  

\begin{enumerate}

\item
We showed that the upcoming {\it GLAST} measurements can be used to
place stringent upper limits, 0.5--5\%, on the ratio of energy
densities of the cosmic rays and thermal gas, $X_{\rm p}$, for several
nearby rich clusters.  These limits are fairly insensitive to the
assumed energy spectrum or the radial distribution of the cosmic-ray
protons for a reasonable range of models.  We showed that IACT
sensitivity to $X_{\rm p}$ is not as stringent as {\it GLAST}, but
IACTs provide useful constraint on spectral index $\alpha_{\rm p}$,
which in turn provide important constraints on the acceleration
mechanisms of cosmic rays.

\vspace{2mm}
\item 
The stacking method offers a powerful technique to probe the
cosmological evolution of $X_{\rm p}$ and $Y_{\rm p}$ with upcoming
$\gamma$-ray observations. Using the latest cosmological models such
as halo mass function and phenomenological relations that reproduce
observed cluster properties, we showed that one-year all-sky survey
with {\it GLAST} can place tight limits ($Y_{\rm p} \lesssim 10\%$) on the
evolution of mean cosmic-ray pressure in clusters out to fairly high
redshift ($z\lesssim 0.25$) by stacking signals from a large sample of
known clusters.  These constraints will correspond to an {\it upper}
limit on the systematic uncertainties in the X-ray and SZE cluster
mass estimates, due to nonthermal pressure provided by cosmic rays.
In addition, since the halo merger rate is expected to increase with
redshift \citep*[e.g.,][]{Gottloeber2001} and such mergers can boost
$\gamma$-ray signals \citep{Pfrommer2007}, the technique may provide
insights into the relation between cosmic-ray energy density and
merger activities.  The same approach will also enable one to probe
cosmic-ray populations in low-mass clusters.

\vspace{2mm}
\item
We also evaluated the cluster contribution to the extragalactic
$\gamma$-ray background using the latest models, and showed that even
with $X_{\rm p} = 1$, the contribution is only about 4\% of the
measured flux.  This indicates that this approach would not currently be
very helpful to constrain $X_{\rm p}$, but might become more useful in
the future if a significant fraction of the background flux were
resolved.

\vspace{2mm}
\item
We showed that $\gamma$-rays due to IC scattering by both the primary
and secondary electrons are likely sub-dominant relative to the
$\gamma$-rays from $\pi^0$ decays in most of the clusters.  We find
that the fractional contribution of the IC flux by secondary electrons
never exceeds $\sim$20\% for a reasonable range of parameters,
independently of $X_{\rm p}$.  The contribution from the primary
electrons will also be suppressed in many clusters, because either
they cool very fast after injection or they cannot be accelerated up
to very high energies in the reacceleration models.  Moreover,
multi-wavelength observations in radio, EUV, and hard X-ray wavebands
will provide independent constraints on nonthermal electrons in
clusters \citep[e.g.,][]{Reimer2004}, and such a consideration shows
that the expected $\gamma$-ray flux from the primary electrons is
indeed sub-dominant as long as $X_{\rm p}>0.02$
(Fig.~\ref{fig:spectrum}).  Even if these components were dominant in
some clusters, the shape of $\gamma$-ray spectrum should provide
diagnostics of the origin.

\end{enumerate}

\section*{Acknowledgments}

We thank Christoph Pfrommer, Julie McEnery, and Steven Ritz for useful
comments. This work was supported by Sherman Fairchild Foundation.

\appendix

\section{\boldmath{$\gamma$}-ray emissivity from \boldmath{$\pi^0$} decays}
\label{app:pion}

Equation~(\ref{eq:emissivity}) has a very clear structure including
several relevant physics, ranging from cosmic-ray distribution $n_{\rm
p}(E_{\rm p},r)$ to $\pi^0$-production cross section $\mathrm
d\sigma_{\rm pp} / \mathrm dE_{\pi}$.
This integral is no difficult, and indeed, \citet{Pfrommer2004} gave a
simple fitting form for that as follows:
\begin{eqnarray}
 q_\gamma (E,r) &=& \sigma_{\rm eff} c n_{\rm H}(r)
 \xi^{2-\alpha_\gamma} \frac{\tilde n_{\rm p}(r)}{\rm GeV}
 \frac{4}{3 \alpha_\gamma}
 \left(\frac{m_\pi c^2}{\rm GeV}\right)^{-\alpha_\gamma}
 \nonumber\\&&{}\times
 \left[\left(\frac{2E}{m_\pi c^2}\right)^{\delta_\gamma}
   +\left(\frac{2E}{m_\pi c^2}\right)^{-\delta_\gamma}
   \right]^{-\alpha_\gamma / \delta_\gamma},
 \nonumber\\
 \label{eq:emissivity 2}
\end{eqnarray}
where $\alpha_\gamma = \alpha_{\rm p}$ is the asymptotic spectral
index of $\gamma$-rays that is the same as that for protons,
$\delta_\gamma = 0.14 \alpha_\gamma^{-1.6} + 0.44$, $\xi = 2$ is a
constant pion multiplicity, and
\begin{equation}
  \sigma_{\rm eff} = 32 (0.96+{\rm e}^{4.4-2.4\alpha_\gamma})
  \, {\rm mb}
\end{equation}
is the effective inelastic pp cross section.
This reproduces results of numerical computations of hadronic
processes as well as accelerator data quite well.

As $\pi^0$'s are produced by collisions between nonthermal cosmic-ray
ions and thermal ICM nucleons, $\gamma$-ray emissivity is proportional
to the product of ICM density $n_{\rm H}$ and number density of cosmic
rays.
The latter quantity is effectively characterized by $\tilde n_{\rm p}$
and this is given by requiring that the fraction $X_{\rm p}$ of the
thermal energy density $\rho_{\rm th}$ goes to cosmic-ray energy
density:
\begin{eqnarray}
  X_{\rm p}(r) \rho_{\rm th}(r)
  & = & \frac{\tilde n_{\rm p}(r) m_{\rm p} c^2}{2 (\alpha_{\rm p}-1)}
  \left(\frac{m_{\rm p}c^2}{\rm GeV}\right)^{1-\alpha_{\rm p}}
  \nonumber\\&&{}\times
  \mathcal B\left(
  \frac{\alpha_{\rm p}-2}{2}, \frac{3-\alpha_{\rm p}}{2}\right),
\end{eqnarray}
where $\mathcal B$ is the beta function, appearing when we integrate
kinetic energy of each proton weighed by the momentum distribution
function, and
\begin{equation}
  \rho_{\rm th}(r) = \frac{3}{2}
  \left(1+\frac{1-3X_{\rm He}/4}{1-X_{\rm He}/2}\right)
  n_{\rm e}(r) k_{\rm B} T_{\rm e}(r),
\end{equation}
with $k_{\rm B}$ the Boltzmann constant, $X_{\rm He} = 0.24$ is the
primordial mass fraction of ${}^4$He, and the electron density $n_{\rm
e}$ and temperature $T_{\rm e}$ are well measured with X-rays.

\section{Inverse-Compton scattering from secondary electrons}
\label{app:IC}

Hadronic collisions also produce charged pions that eventually decay
into electrons and positrons.
These leptons, having relativistic energies, can up-scatter the CMB
photons into GeV energies.
Since the physics of IC scattering is well established
\citep{Rybicki1979} and pion production due to pp collisions are
measured in laboratories, this process can be described with
relatively small ambiguity.

Electron distribution function after radiative cooling is
\begin{equation}
  n_{\rm e}(E_{\rm e},r) = \frac{\tilde n_{\rm e}(r)}{\rm GeV}
  \left(\frac{E_{\rm e}}{\rm GeV}\right)^{-\alpha_{\rm e}},
\end{equation}
\begin{equation}
  \tilde n_{\rm e}(r) =
  \frac{2^7\pi 16^{-(\alpha_{\rm e}-1)}}{\alpha_{\rm e}-2}
  \frac{\sigma_{\rm eff}m_{\rm e}^2c^4}{\sigma_{\rm T}\,{\rm GeV}}
  \frac{n_{\rm H}(r)\tilde n_{\rm p}(r)}{B(r)^2+B_{\rm CMB}^2},
\end{equation}
where $\alpha_{\rm e} = \alpha_{\rm p} + 1$, $\sigma_{\rm T}$ is the
Thomson cross section, and $B_{\rm CMB} = 3.24(1+z)^2$ $\mu$G.
Emissivity of IC scattered photons is given as
\begin{eqnarray}
  q_{\rm IC}(E,r) & = & \tilde q(r) f_{\rm IC}(\alpha_{\rm e})
  \left(\frac{m_{\rm e}c^2}{\rm GeV}\right)^{1-\alpha_{\rm e}}
  \nonumber\\&&{}\times
  \left(\frac{E}{k_{\rm B}T_{\rm
  CMB}}\right)^{-(\alpha_\nu+1)},\\
  \tilde q(r) &=& \frac{8\pi^2 r_{\rm e}^2 \tilde n_{\rm e}(r)
    (k_{\rm B}T_{\rm CMB})^2}{h^3c^2},\\
  f_{\rm IC}(\alpha_{\rm e}) &=& \frac{2^{\alpha_{\rm e}+3}
    (\alpha_{\rm e}^2+4\alpha_{\rm e}+11)}{(\alpha_{\rm e} + 3)^2
  (\alpha_{\rm e}+5)(\alpha_{\rm e}+1)}
  \nonumber\\&&{}\times
  \Gamma\left(\frac{\alpha_{\rm e}+5}{2}\right)
  \zeta\left(\frac{\alpha_{\rm e}+5}{2}\right),
\end{eqnarray}
where $\alpha_\nu = (\alpha_{\rm e} - 1)/2$, $r_{\rm e}$ is the
classical electron radius, $T_{\rm CMB} = 2.7$ K is the CMB
temperature, $\Gamma$ is the $\Gamma$-function, and $\zeta$ is the
Riemann $\zeta$-function.
For more detailed discussions, see \citet{Dolag2000} and
\citet{Pfrommer2004}.

\section{\textit{GLAST} sensitivity for extended sources}
\label{app:sensitivity}

Flux sensitivity of {\it GLAST}-LAT to a point-like source is shown as a
thick dashed line in Fig.~\ref{fig:spectrum}.
The sensitivity for an extended source is different; in this section, we
derive it using a simple argument.

The dominant background is extragalactic $\gamma$-ray flux, and its
intensity depends on photon energy as $I_{\rm bg} \propto E^{-2.1}$
\citep{Sreekumar1998}.
At low energies where background photon count are larger than 1 ($N_{\rm
bg} \gtrsim 1$), the flux sensitivity is determined by a criterion such
as $N_{\rm lim} > \alpha \sqrt{N_{\rm bg}}$, where some number $\alpha$
sets significance of detection; hereafter use $\alpha = 5$.
On the other hand, at higher energies where $N_{\rm bg} \lesssim 1$,
then the detection simply relies upon photon count from a source.

Thus, for a source like galaxy clusters, there are up to four different
energy regimes depending on the source extention.
(i) At lowest energies where PSF (or pixel size) is larger than the
source size (i.e., $\Omega_{\rm pix} > \Omega$), the source can be
regarded as a point-like object.
The other three regimes are for more energetic photons where source are
extended ($\Omega > \Omega_{\rm pix}$); they are where background photon
counts are (ii) larger than 1 from one pixel ($N_{\rm bg,pix} > 1$);
(iii) smaller than 1 from one pixel but larger than 1 from the entire
source region ($N_{\rm bg,pix} < 1, N_{\rm bg} > 1$); (iv) smaller than
1 from the entire source region ($N_{\rm bg} < 1$).
For the lowest energy region (i), the sensitivity is the same as that
for point sources, and this corresponds to the regime below $\sim$300
MeV in Fig.~\ref{fig:spectrum}.

For the region (ii), the point-source flux sensitivity $F_{\rm lim,pix}$
are obtained by a criterion $N_{\rm lim,pix} = 5 \sqrt{N_{\rm
bg,pix}}$.
These photon numbers are related to the flux and background intensity
through $N_{\rm lim, pix} = F_{\rm lim, pix} A T$ and $N_{\rm bg, pix} =
I_{\rm bg} A T \Omega_{\rm pix}$, where $A T$ (effective area times
exposure time) is the detector exposure.
The point-source sensitivity is thus obtained by
\begin{equation}
 F_{\rm lim,pix} = 5 \sqrt{\frac{I_{\rm bg} \Omega_{\rm pix}}{AT}}.
\end{equation}
A similar argument can be applied for the flux sensitivity for an
extended source $F_{\rm lim}$ and we obtain
\begin{equation}
 F_{\rm lim} = 5 \sqrt{\frac{I_{\rm bg} \Omega}{AT}}
  = F_{\rm lim,pix} \sqrt{\frac{\Omega}{\Omega_{\rm pix}}}.
  \label{eq:F_lim}
\end{equation}
Thus the sensitivity becomes weaker by a factor of $\theta / \delta
\theta_{\rm PSF}$, compared to that for a point-like source.
This is the case for the region between 300 MeV and 2 GeV in
Fig.~\ref{fig:spectrum}.

When the photon energy becomes higher, the background count gets
smaller.
We then consider the region (iii).
In this case, to obtain the point-source sensitivity, we use a criterion
of five-photon detection: $F_{\rm lim,pix} = 5/AT$.
As we have more background photons (than 1) from the entire source
region, the extended-source flux sensitivity is the same as the first
equality of equation~(\ref{eq:F_lim}).
Combining these two, we obtain
\begin{equation}
 F_{\rm lim} = \sqrt{5 F_{\rm lim,pix} I_{\rm bg} \Omega},
\end{equation}
and this is for the region between 2 GeV and 30 GeV in
Fig.~\ref{fig:spectrum}.
At the highest regime (iv), the region above 30 GeV in
Fig.~\ref{fig:spectrum}, the cluster detection is totally relies on
photon count and independent of background.
Therefore, the sensitivity for an extended source is the same as that
for a point-like source.

\label{lastpage}

\end{document}